\documentclass[preprint,aps,superscriptaddress]{revtex4}
\usepackage{graphicx}
\usepackage{dcolumn}
\usepackage{bm}
\usepackage{times}

\begin{document}

\title{Fast light, slow light, and phase singularities: a connection to
generalized weak values}
\author{D. R. Solli}
\affiliation{Department of Physics, University of California, Berkeley, CA 94720-7300.}
\author{C. F. McCormick}
\affiliation{Department of Physics, University of California, Berkeley, CA 94720-7300.}
\author{R. Y. Chiao}
\affiliation{Department of Physics, University of California, Berkeley, CA 94720-7300.}
\author{S. Popescu}
\affiliation{Univ Bristol, HH Wills Phys Lab, Bristol BS8 1TL, Avon, England.}
\affiliation{Hewlett Packard Labs, Bristol BS12 6QZ, Avon, England.}
\author{J. M. Hickmann}
\affiliation{Department of Physics, University of California, Berkeley, CA 94720-7300.}
\affiliation{Departamento de F\'{\i}sica, Universidade Federal de Alagoas, Cidade
Universit\'{a}ria, 57072-970, Macei\'{o}, AL, Brazil.}

\begin{abstract}
We demonstrate that Aharonov-Albert-Vaidman (AAV) weak values have a direct
relationship with the response function of a system, and have a much wider
range of applicability in both the classical and quantum domains than
previously thought. Using this idea, we have built an optical system, based
on a birefringent photonic crystal, with an infinite number of weak values.
In this system, the propagation speed of a polarized light pulse displays
both superluminal and slow light behavior with a sharp transition between
the two regimes. We show that this system's response possesses
two-dimensional, vortex-antivortex phase singularities. Important
consequences for optical signal processing are discussed.
\end{abstract}

\maketitle

The role of measurement in quantum mechanics has posed challenges for
physicists for nearly a century. Standard quantum mechanics represents ideal
measurements as the action of a Hilbert-space operator on a state. In an
ideal or \textquotedblleft strong\textquotedblright\ measurement, the
operator completely collapses the state into an eigenstate, with an
expectation value within the range of the operator's eigenvalues. Strong
measurements yield maximal information about the state of the system and
correspond to most laboratory measurements.

Contrary to traditional quantum measurement theory, certain experiments can
return results that are far beyond the range of an observable's eigenvalues 
\cite{Aharonov1988}. These so-called \textquotedblleft
weak\textquotedblright\ measurements barely perturb the state, and have been
less studied in quantum mechanics because they yield very little
information. In their seminal work, Aharonov, Albert and Vaidman (AAV)
showed that paradoxical results can be obtained if one preselects an initial
state, performs a weak measurement, and subsequently a standard strong
measurement which postselects a particular final state. Although the action
of the weak measurement has little effect on the global state of the system,
it can have a significant impact on one or more of its components. Weak
values have been theoretically discussed for spin-$1/2$ particles and
experimentally demonstrated in optical systems involving the angular
deflections of polarized light beams passing through birefringent prisms 
\cite{Aharonov1988,Duck1989,Knight1990,Ritchie1991}.

Recently, using the approach of quantum-trajectory theory, it has been demonstrated that postselected weak values can be thought of as quantum correlation functions, which can be used in quantum optics and related areas \cite{Wiseman2002}. In this paper, we instead start from the original derivation of the AAV effect and extend it further. 

We show that weak values appear in a natural way in a much
broader range of situations. We demonstrate that a system's response
function (or scattering matrix) has a direct relationship with AAV weak
values. As an illustration, we present an experimental study of a
frequency-dependent, polarization-sensitive, optical system based on a
two-dimensional (2D), birefringent photonic crystal. The response of this
system can be described by an infinite number of unbounded weak values which
physically correspond to measurements of the polarization value or spin of
the photon along different axes. Unlike previously studied weak
measurements, this system maps polarization states onto collinear spatial
position along the direction of propagation, resulting in unbounded values
for the time-of-flight of a polarized optical pulse. The system displays
both superluminal \cite{Steinberg1993,Wang2000} and slow light \cite{Hau1999}%
\ behavior with a sharp transition between the two regimes and gives us
total control over the speed of light. It may lead to several important
applications in optical signal processing \cite{Lukin2001}, such as optical
delay lines with polarization-tuned characteristics and dynamical optical
clock multipliers. This new approach to quantum weak values should also open
the door for many new precision measurements of very tiny quantities \cite%
{Aharonov1988,Duck1989}.

The complex response of a linear system parametrized by two general
variables $\left( \rho ,\eta \right) $ is given by 
\begin{equation}
\tilde{T}\left( \rho ,\eta \right) =\left\langle \psi _{f}\right\vert
U\left( \rho ,\eta \right) \left\vert \psi _{in}\right\rangle
\label{transfer}
\end{equation}

\noindent where $U\left( \rho ,\eta \right) $ is a unitary operator
describing the evolution of the system, and $\left\vert \psi
_{in}\right\rangle $ and $\left\vert \psi _{f}\right\rangle $ are normalized
input and output states. To apply the AAV formalism, we must express Eq. \ref%
{transfer} in terms of the preselection of an initial state, a weak
measurement, and a strong measurement or projection onto a particular final
state. The difficulty lies in the fact that $U\left( \rho ,\eta \right) $
cannot be generally classified as a weak operator because it may have a
strong impact on $\left\vert \psi _{in}\right\rangle $. However, we are free
to decompose $U\left( \rho ,\eta \right) $ into a large or \textquotedblleft
strong\textquotedblright\ part and a small or \textquotedblleft
weak\textquotedblright\ part. To do so, we Taylor expand the general
operator $U\left( \rho ,\eta \right) $ around an arbitrary point $\left(
\rho _{0},\eta _{0}\right) $ for small deviations $\delta \rho $ and $\delta
\eta $, and separate $U\left( \rho ,\eta \right) $ into a weak operator $U_{%
\mathrm{W}}\left( \delta \rho ,\delta \eta ;\rho _{0},\eta _{0}\right) $ and
a strong operator $U_{\mathrm{S}}\equiv U\left( \rho _{0},\eta _{0}\right) $
obtaining 
\[
\tilde{T}\left( \rho _{0}+\delta \rho ,\eta _{0}+\delta \eta \right) \approx 
\]%
\begin{equation}
\left\langle \psi _{f}\right. \left\vert \tilde{\psi}_{in}\left( \rho
_{0},\eta _{0}\right) \right\rangle \left\{ 1+i\left[ A_{\rho }\right]
_{W}\delta \rho +i\left[ A_{\eta }\right] _{W}\delta \eta \right\}
\label{transfer-approx}
\end{equation}

\noindent where $\left\vert \tilde{\psi}_{in}\left( \rho _{0},\eta
_{0}\right) \right\rangle \equiv U_{\mathrm{S}}\left\vert \psi
_{in}\right\rangle $, and $\left[ A_{\rho }\right] _{W}$ and $\left[ A_{\eta
}\right] _{W}$ are given by: 
\begin{equation}
\left[ A_{j}\left( \rho _{0},\eta _{0}\right) \right] _{W}=\frac{%
\left\langle \psi _{f}\right\vert A_{j}\left( \rho _{0},\eta _{0}\right)
\left\vert \tilde{\psi}_{in}\left( \rho _{0},\eta _{0}\right) \right\rangle 
}{\left\langle \psi _{f}\right. \left\vert \tilde{\psi}_{in}\left( \rho
_{0},\eta _{0}\right) \right\rangle }  \label{weak}
\end{equation}%
\noindent with $A_{j}\left( \rho _{0},\eta _{0}\right) =-i\left[ \partial
_{j}U_{\mathrm{S}}\right] U_{\mathrm{S}}^{\dagger }$ for $j\rightarrow \rho
,\eta $.

The strong operator $U_{\mathrm{S}}$ can be viewed as a preselector which
generates the state $\left\vert \tilde{\psi}_{in}\left( \rho _{0},\eta
_{0}\right) \right\rangle $ from $\left\vert \psi _{in}\right\rangle $, and $%
U_{\mathrm{W}}\left( \delta \rho ,\delta \eta ;\rho _{0},\eta _{0}\right) $
is a weak measurement operator which acts on this preselected state. Eq. \ref%
{weak} then implies that $\left[ A_{\rho }\right] _{W}$ and $\left[ A_{\eta }%
\right] _{W}$ are AAV weak values associated with the operators $A_{\rho
}\left( \rho _{0},\eta _{0}\right) $ and $A_{\eta }\left( \rho _{0},\eta
_{0}\right) $. These weak values $\left[ A_{\rho }\right] _{W}$ and $\left[
A_{\eta }\right] _{W}$ can be unbounded if the corresponding transfer
function has a singularity, i.e., a point for which $\left\langle \psi
_{f}\right. \left\vert \tilde{\psi}_{in}\left( \rho _{0},\eta _{0}\right)
\right\rangle =0$.

\begin{figure}
\centerline{\includegraphics{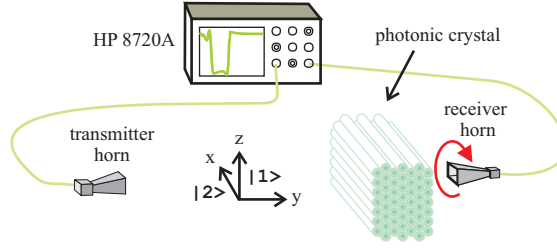}}
\caption{Experimental setup for transmission measurements.}
\label{newsetup}
\end{figure}

\begin{figure}[h]
\centerline{\includegraphics{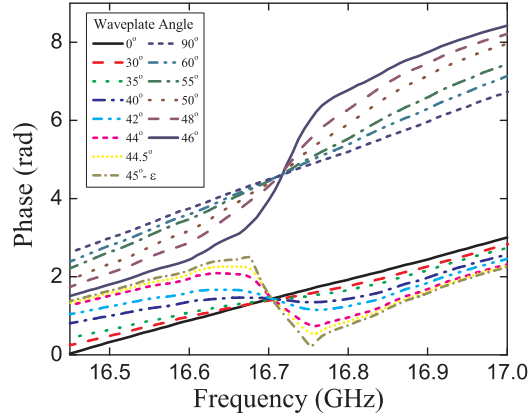}}
\caption{Experimentally measured phase delay vs. frequency for microwaves
transmitted through an 18-layer photonic crystal positioned at different
angles.}
\label{data}
\end{figure}

\begin{figure}[h]
\centerline{\includegraphics{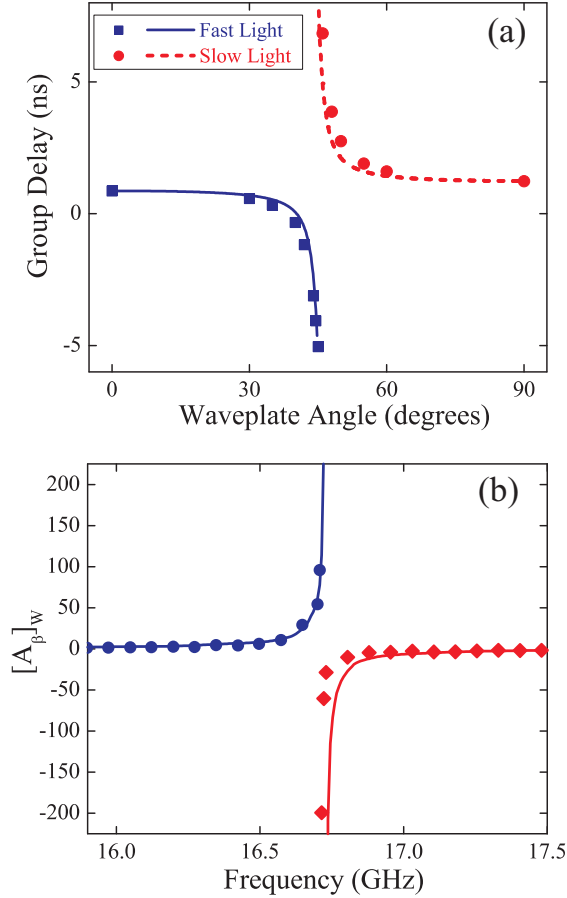}}
\caption{Experimentally measured group delay vs. crystal rotation at $%
\protect\omega $ = 16.7 GHz (a) and pointer for $\left[ A_{\protect\beta }%
\right] _{W}$ vs. frequency at $\protect\beta =\protect\pi /4$ (b). }
\label{delay}
\end{figure}

\begin{figure}[h]
\centerline{\includegraphics{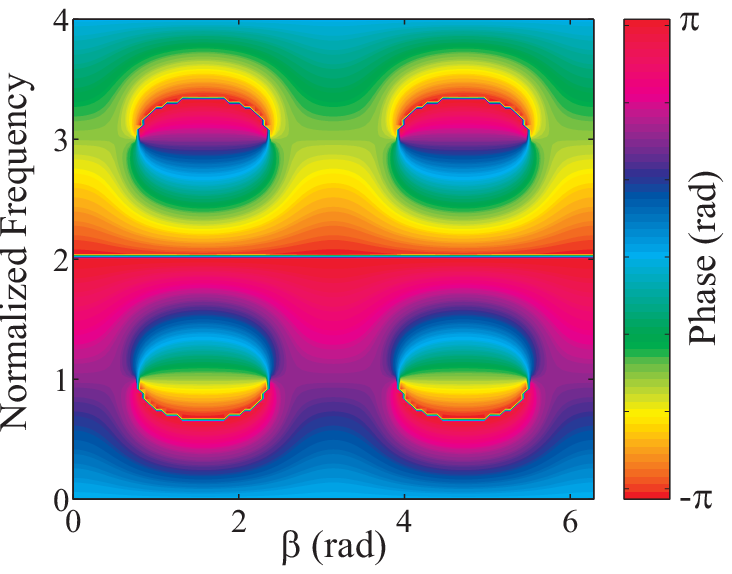}}
\caption{Theoretical contour plot of $\left[ \arg \tilde{T}(\protect\omega ,%
\protect\beta )\right] $ showing the singularities and the surrounding
phase. The arcs connecting the singularities correspond to $2\protect\pi $
phase slips.}
\label{contours}
\end{figure}

This analysis is valid for classical and quantum-mechanical systems. In the
quantum mechanical case, we must reinterpret $\delta \rho $ and $\delta \eta 
$ as operators rather than deviation parameters, while, the preselected
state is a superposition of eigenstates, sharply peaked about $\rho _{0}$
and $\eta _{0}$ such that the expectation values of $\delta \rho $ and $%
\delta \eta $, appearing in Eq. \ref{transfer-approx}, are small relative to 
$\rho _{0}$ and $\eta _{0}$, respectively.

Experimentally observable quantities are inferred from an indicator or
pointer on a measuring device. In the general case, $\delta \rho $ and $%
\delta \eta $ are canonical variables describing two independent measuring
devices. The pointers of these measuring devices are the conjugate momenta
to these canonical variables, and the interaction weak operator $U_{W}$
couples them to the physical observables $A_{\rho }$ and $A_{\eta }$. As a
result, the measurement process shifts the pointer by an amount equal to the
weak value. Our analysis shows that there is always a well-defined
connection between measurement pointers for weak values and the response
function of a system. To make this explicit, we invert Eq. \ref%
{transfer-approx}, obtaining 
\[
\left[ A_{j}\left( \rho _{0},\eta _{0}\right) \right] _{W}= 
\]%
\begin{equation}
\partial _{j}\left[ \arg \tilde{T}\left( \rho _{0},\eta _{0}\right) \right]
-i\partial _{j}\left[ \ln \left\vert \tilde{T}\left( \rho _{0},\eta
_{0}\right) \right\vert \right] \text{.}  \label{inv-weak}
\end{equation}%
\noindent In words, the local gradients of the phase and the logarithm of
the magnitude of a response function are measurement pointers for the real
and imaginary parts of weak values.

To illustrate this formalism we apply it to experimental measurements of the
transmission of normally incident, linearly polarized microwaves through a
photonic crystal. These crystals, constructed using a previously published
method \cite{Hickmann2002}, are highly birefringent \cite%
{Solli2003-1,Solli2003-2}. Our crystal can be rotated in the $xz$-plane by
an angle $\beta $ with respect to the $z$-axis and its fast axis is parallel
to $\hat{x}$ at $\beta =0$. Microwaves polarized along \ $\widehat{z}$ are
generated by a HP 8720A vector network analyzer (VNA) and detected by a
polarization-sensitive horn (see Fig. 1). The two degrees of freedom $\omega 
$ and $\beta $ allow the transmitted polarization to lie anywhere on the 2D
Hilbert space of polarizations (Poincar\'{e} sphere). If the crystal is
positioned at $\beta =\pi /4$, there is no detected signal at $\omega _{%
\mathrm{s}}$, the frequency for which the crystal behaves as a half
waveplate, because the transmitted EM wave is entirely $\hat{x}$-polarized.

In Fig. 2, we display experimental results for the phase delay as a function
of frequency for microwaves transmitted through an 18-layer photonic crystal
($\omega _{\mathrm{s}}=$ 16.7 GHz) positioned at various angles. The phase
delay exhibits strong dispersion near $\beta =\pi /4$. Surprisingly, for $%
\beta <\pi /4$ the dispersion is anomalous, whereas it is normal for $\beta
>\pi /4$. Some insight into this behavior can be found from a general
topological argument. If the parameters $\left( \omega ,\beta \right) $ are
swept to trace out a closed loop around the singularity (zero-transmission
point), the phase can change by an integer multiple of $2\pi $ over a
complete cycle. The phase may therefore show extremely rapid and even
discontinuous changes as these parameters are adjusted through paths that
pass close to or over the singularity.

To apply our framework to the experiment, we construct the unitary operator $%
U\left( \omega ,\beta \right) $ in terms of the usual Pauli matrices $\sigma
_{k}$, finding 
\begin{equation}
U(\omega ,0)=\exp [i\phi _{\mathrm{TE}}(\frac{1+\sigma _{3}}{2})+i\phi _{%
\mathrm{TM}}(\frac{1-\sigma _{3}}{2})]
\end{equation}

\noindent where $\phi _{\mathrm{TE}},\ \phi _{\mathrm{TM}}$ are the
birefringent phases for TE and TM polarizations and $k\rightarrow 1,2,3$.
The independent vertical and horizontal polarization states $\left\vert
1\right\rangle $ and $\left\vert 2\right\rangle $ are parallel to $\hat{z}$
and $\hat{x}$, respectively, and $\sigma _{3}\left\vert 1\right\rangle
=-\left\vert 1\right\rangle $ and $\sigma _{3}\left\vert 2\right\rangle
=\left\vert 2\right\rangle $. $U(\omega ,\beta )$ is found by a simple
rotation, and the corresponding weak operators follow naturally from the
Taylor expansion.

Our weak values are associated with polarization measurements of the
incident radiation. For our experimental system, the imaginary part of Eq. %
\ref{inv-weak} ($j\rightarrow \omega $) vanishes for points along the line $%
\left( \omega _{\mathrm{s}},\beta _{0}\right) $. Thus along the line $%
(\omega _{\mathrm{s}},\beta _{0})$ the pointer for $\left[ A_{\omega }\right]
_{W}$ is simply the frequency derivative of the transmitted phase, or the
group delay of a pulse with narrow spectral width $\delta \omega $ and
center frequency $\omega _{\mathrm{s}}$. The weak value $\left[ A_{\omega }%
\right] _{W}$ is coupled to a spatial position pointer and manifested as a
time-of-flight. A calculation of the weak value yields%
\[
\lbrack A_{\omega }(\omega _{s},\beta _{0})]_{W}=\sec (2\beta _{0})[\cos
^{2}(\beta _{0})\partial _{\omega }\phi _{\mathrm{TM}}(\omega _{\mathrm{s}}) 
\]%
\begin{equation}
-\sin ^{2}(\beta _{0})\partial _{\omega }\phi _{\mathrm{TE}}(\omega _{%
\mathrm{s}})]\text{.}  \label{group-delay}
\end{equation}

By taking numerical derivatives of the measured phase delay with respect to
the frequency, we generate experimental data for the pointer for $\left[
A_{\omega }\left( \omega _{s},\beta _{0}\right) \right] _{W}$. In Fig. 3(a)
we show the results at $\omega _{\mathrm{s}}$ as a function of the waveplate
angle. For comparison, we also show theoretical curves generated using Eq. %
\ref{group-delay} and independent measurements of $\phi _{TM}\left( \omega
\right) $ and $\phi _{TE}\left( \omega \right) $.

The group delay measurement pointer shows both \textquotedblleft
fast\textquotedblright\ and \textquotedblleft slow\textquotedblright\ light
behavior; on one side of the singularity at $\beta =\pi /4$, arbitrarily
large, negative group velocities are possible ($\beta <\pi /4$), while on
the other side ($\beta >\pi /4$), pulses can propagate with arbitrarily
small group velocities. The pointer takes different positive, non-degenerate
values at $\beta =0$ and $\beta =\pi /2$. As a result, this passive system
can be used as a compact optical delay line whose properties can be tuned by
small adjustments of the waveplate angle and the birefringence. Since the
birefringent characteristics of a photonic crystal can be tailored by
changing its geometry, air-filling fraction, and material composition \cite%
{Solli2003-2}, this type of delay line can be adapted to suit a wide variety
of applications.

The weak value $\left[ A_{\beta }\left( \omega _{0},\beta _{s}\right) \right]
_{W}$ is also a polarization value or spin of the photon. Near the
singularity, the complicated form\ of the operator $A_{\beta }\left( \omega
_{0},\beta _{s}\right) $ reduces to $2\sigma _{2}\equiv L_{y}/\hbar $ to
leading order in $\omega _{0}$. Thus, in this vicinity, the weak value is
simply the total helicity of the light about the $y$-axis. Along the line $%
\left( \omega _{0},\beta _{s}\right) $ where $\beta _{s}=\pi /4$, the
pointer for the weak value reduces to the real quantity $\partial _{\beta }%
\left[ \arg \tilde{T}\left( \omega _{0},\beta _{s}\right) \right] $. An
explicit calculation of the pointer on this line produces the simple form%
\begin{equation}
\left[ A_{\beta }\left( \omega _{0},\beta _{s}\right) \right] _{W}=2\tan %
\left[ \phi _{-}\left( \omega _{0}\right) \right] \text{.}  \label{angular}
\end{equation}%
If $\phi _{-}\left( \omega _{0}\right) =\frac{1}{2}\left[ \phi _{TE}(\omega
)-\phi _{TM}(\omega )\right] =0$ or $\pi $, Eq. \ref{angular} yields the
degenerate eigenvalues $\left[ A_{\beta }\left( \omega _{0},\beta
_{s}\right) \right] _{W}=0$; however, the pointer assumes arbitrarily large
positive and negative values near the tangent singularity, where it
corresponds to a measurement of the total helicity of the light about the $y$%
-axis. Since the number of photons in light is of course finite, this
unbounded weak value would imply that the basic unit of angular momentum
carried by each photon can be anything. In Fig. 3(b), we display
experimental data for this pointer as a function of frequency with $\beta
=\pi /4$.

One may wonder why the crystal itself cannot be used as a measurement
pointer for this weak value since the angular momentum of the crystal about
the $y$-axis is conjugate to $\beta $. As a classical object, it cannot
exist in a superposition of angular states. Therefore, its angular momentum
is uncertain on the quantum scale and will not reflect the weak value. This
point emphasizes the value of Eq. \ref{inv-weak}; local derivatives of the
response function can always be used as measurement pointers for weak
values, regardless of whether the experiment is classical or
quantum-mechanical. Similar reasoning applies to the group delay pointer for 
$\left[ A_{\omega }\right] _{W}$, since a pulse is simply a classical
superposition of different frequencies: if the weak measurement is performed
with continuous-wave light, the time-of-flight is not a well-defined
quantity, but can be directly inferred from $\partial _{\omega }\left[ \arg 
\tilde{T}\left( \omega _{0},\beta _{0}\right) \right] $.

In Fig. 4, we display a theoretical contour plot of $\arg \tilde{T}\left(
\omega ,\beta \right) $ generated using simple linear models for $\phi
_{TM}\left( \omega \right) $ and $\phi _{TE}\left( \omega \right) $.
Singularities are expected at all points $\left( \omega ^{n},\beta
^{m}\right) $ where $\phi _{-}\left( \omega ^{n}\right) =\left( 2n+1\right) 
\frac{\pi }{2}$ and $\beta ^{m}=\left( 2m+1\right) \frac{\pi }{4}$ for
integers $n$ and $m$. Using these models, we display eight singularities,
and the topology of the surrounding phase. The real parts of the weak values 
$\left[ A_{\omega }\left( \omega _{0},\beta _{0}\right) \right] _{W}$ and $%
\left[ A_{\beta }\left( \omega _{0},\beta _{0}\right) \right] _{W}$ at any
point are given by the local gradient of the phase. Between the
singularities, there are saddle points, i.e., points where the weak values
are least sensitive to deviations in the parameters. The global structure is
that of a vortex-antivortex lattice, with adjacent singularities possessing
opposite topological charges. This structure implies that the system is
relatively insensitive to perturbations in the operator $U\left( \omega
,\beta \right) $ because the spontaneous appearance or disappearance of a
singularity would have a discontinuous effect on the global phase. Thus, the
topology must possess an equal number of vortices and antivortices; as the
unitary operator is continuously perturbed, vortices and antivortices can
either approach each other and annihilate or separate to infinity \cite%
{Berry2003}. Finally we note that while $\left[ A_{\omega }\right] _{W}$ and 
$\left[ A_{\beta }\right] _{W}$ are the natural weak values of this system,
arbitrary superpositions of the two weak operators lead to infinite number
of other weak values whose physical interpretation is less clear.

In conclusion, we have shown that response functions can be treated with the
AAV formalism, although the physical interpretation is specific to the
particular physical system under study. If the response possesses
singularities, one obtains weak values which can be far outside the ranges
spanned by the eigenvalues of the corresponding operators. We have
demonstrated this using an optical system exhibiting singular points in a
two-dimensional state space characterized by two independent weak values and
an infinite number of superpositions thereof. This system has significant
implications for optical communication applications requiring compact,
tunable delay lines and clock multipliers. Although our experiments were
performed using classical signals, the results apply to quantum-mechanical
experiments as well.

This work was supported by ARO, grant number DAAD19-02-1-0276, ONR and NSF.
We thank the UC Berkeley Astronomy Department, in particular Dr. R.
Plambeck, for lending us the VNA. JMH thanks the support from Instituto do
Mil\^{e}nio de Informa\c{c}\~{a}o Qu\^{a}ntica, CAPES, CNPq, FAPEAL,
PRONEX-NEON, ANP-CTPETRO.

\end{document}